\documentclass[12pt]{article}
\usepackage{graphicx}
\usepackage[cp1251]{inputenc}

\begin{document}
 \noindent {\footnotesize\it
   Astronomy Letters, 2021, Vol. 47, No 3, pp. 180--187.}
 \newcommand{\dif}{\textrm{d}}

 \noindent
 \begin{tabular}{llllllllllllllllllllllllllllllllllllllllllllll}
 & & & & & & & & & & & & & & & & & & & & & & & & & & & & & & & & & & & & & &\\\hline\hline
 \end{tabular}

  \vskip 0.5cm
  \centerline{\bf\large Study of Close Stellar Encounters with the Solar System Based on}  \centerline{\bf\large Data from the Gaia EDR3 Catalogue}
   \bigskip
   \bigskip
  \centerline
 {V.V. Bobylev\footnote [1]{e-mail: vbobylev@gaoran.ru}  and A. T. Bajkova}
   \bigskip

  \centerline{\small\it Pulkovo Astronomical Observatory, Russian Academy of Sciences,}

  \centerline{\small\it Pulkovskoe sh. 65, St. Petersburg, 196140 Russia}
 \bigskip
 \bigskip
 \bigskip

 {
{\bf Abstract}---We have studied stellar candidates for close (within 1 pc) encounters with the Solar system. For all of the stars under consideration the kinematic characteristics have been taken from the Gaia EDR3 catalogue. The parameters of the encounters of these stars with the Solar system have been calculated using three methods: (1) the linear one, (2) by integrating the orbits in an axisymmetric potential, and (3) by integrating the orbits in a potential with a spiral density wave. All three methods are shown to yield similar results. We have selected five stars that are good candidates for reaching the boundaries of the Oort cloud and passing through it. Based on the second method, in good agreement with the other two methods, we have obtained the following estimates of the encounter parameters for the star GJ 710:
$t_{min}=1.320\pm0.028$ Myr and $d_{min}=0.020\pm0.007$ pc. It is also interesting to note the star Gaia EDR3 510911618569239040 with parameters $t_{min}=-2.863\pm0.046$ Myr and $d_{min}=0.057\pm0.079$ pc.
  }

 % DOI: 10.1134/S1063773721020031

 \subsection*{INTRODUCTION}
Close (within 1--2 pc) encounters of field stars with the Solar system can lead to a perturbation of the Oort cloud (Oort 1950). Such a perturbation can provoke the emergence of the so-called comet shower from the outer boundaries of the Oort cloud into the inner Solar system, in particular, toward the Earth. As simulations show (Dybczy\'nski 2002, 2005;
Martinez-Barbosa et al. 2017), apart from stellar flybys, the Oort comet cloud is subject to perturbations from giant molecular clouds and experiences an effect from a Galactic tide. According to Oort (1950), the presumed outer boundary of the Oort cloud is
$\sim10^5$ AU (0.48 pc).

A new surge of interest in the evolutionary properties of the Oort cloud is related to the detection of two interstellar wanderers in the Solar system, namely 1I/’Oumuamua (Bacci et al. 2017) and 2I/Borisov (Borisov 2019). According to the estimates by Portegies
Zwart (2020), $\sim$6\% of the nearest stars may have planets and asteroids in their Oort clouds. Such bodies can be liberated from the parent star and escape into interstellar space. Moving in a Galactic orbit close to the orbit of the parent star, these bodies
form dense streams of rogue interstellar asteroids and planets. The Solar system occasionally passes through such streams, potentially giving rise to occasional
close encounters with objects from these streams.

A practical search for close stellar encounters with the Solar system was carried out in the pioneering studies by Revina (1988), Matthews (1994), and M\"ull\"ari and Orlov (1996). These authors revealed a number of interesting candidates, for example, Proxima
Centauri, the $\alpha$~Centauri system, or the star GJ 905.

Based on data from the Hipparcos (1997) catalogue, this problem was solved by Garcia-S\'anchez et al. (1999, 2001), Bobylev (2010a, 2010b), Anderson
and Francis (2012), Dybczy\'nski and Berski (2015),
Bailer-Jones (2015), and Feng and Bailer-Jones (2015).
The search for close encounters based on data
from the Gaia TGAS (Tycho–Gaia Astrometric
Solution, Lindegren et al. 20016) catalogue led
to the detection of several candidates for a very
close flyby (Berski and Dybczy\'nski 2016; Bobylev
and Bajkova 2017; R. de la Fuente Marcos and C. de la Fuente Marcos 2018), namely for an injection into the Oort cloud (to distances less than 0.5 pc).
One of the record-holders is the star GJ 710 (Garcia-S\'anchez et al. 2001; Bobylev 2010a; Berski and Dybczy\'nski 2016; Bailer-Jones 2018).

An analysis of the Gaia DR2 data (Brown et al. 2018; Lindegren et al. 2018) showed (see Bailer-Jones et al. 2018; Darma et al. 2019; Torres et al. 2019; Wysocza\'nska et al. 2020; Bobylev and Bajkova 2020) that $\sim$3000 candidates, $\sim$30 stars, and 5--6 stars
can have encounters with the Solar system within 5, 1, and 0.25 pc, respectively, on a time interval of $\pm$5 Myr.

In the latest Gaia EDR3 (Gaia Early Data Release 3, Brown et al. 2020; Lindegren et al. 2020) version the trigonometric parallaxes and proper motions
were improved approximately by 30\% for $\sim$1.5 billion
stars. In contrast, the line-of-sight velocities were
just copied from the Gaia DR2 catalogue. Therefore,
the Gaia EDR3 data in the search for close encounters can efficiently serve to improve the encounter parameters of the already revealed candidates.

The goal of our paper is the application of various methods of analyzing the motion of the candidate stars to improve the parameters of their close encounters
with the Solar system using the latest measurements
of stellar parallaxes and proper motions from
the Gaia EDR3 catalogue. The linear method and
two Galactic potentials, an axisymmetric one and a nonaxisymmetric one including a spiral density wave, are used to construct the stellar orbits.

 \subsection*{ORBIT CONSTRUCTION METHODS}
In a rectangular coordinate system with the center in the Sun the $X$ axis is directed toward the Galactic center, the $Y$ axis is in the direction of Galactic rotation,
and the $Z$ axis is directed to the north Galactic pole.
Then, $X=r\cos l\cos b,$ $Y=r\sin l\cos b,$ and $Z=r\sin b,$ where $r=1/\pi$ is the stellar heliocentric distance in kpc that we calculate via the stellar parallax
$\pi$ in mas. Note that in this paper we use stars with relative parallax errors less than 10\% and, therefore, there is no need to take into account the Lutz–Kelker
bias (Lutz and Kelker 1973).

The line-of-sight velocity $V_r$ and the two tangential velocity components 
$V_l=4.74r\mu_l\cos b$ and $V_b=4.74r\mu_b$ along the Galactic longitude $l$ and latitude $b,$ respectively, expressed in km s$^{-1}$ are known from observations. Here, the coefficient 4.74 is the ratio of the number of kilometers in an astronomical unit to
the number of seconds in a tropical year. The proper motion components $\mu_l\cos b$ and $\mu_b$ are expressed in mas yr$^{-1}$.

The velocities $U, V,$ and $W,$ where $U$ is directed away from the Sun toward the Galactic center, $V$ is in the direction of Galactic rotation, and $W$ is directed to the north Galactic pole, are calculated via the components $V_r, V_l,$ and $V_b,$ respectively:
 \begin{equation}
 \begin{array}{lll}
 U=V_r\cos l\cos b-V_l\sin l-V_b\cos l\sin b,\\
 V=V_r\sin l\cos b+V_l\cos l-V_b\sin l\sin b,\\
 W=V_r\sin b                +V_b\cos b.
 \label{UVW}
 \end{array}
 \end{equation}

 \subsubsection*{\it Linear Method}
According to Matthews (1994), the minimum distance between the stellar and solar trajectories $d_{min}$ at the encounter time $t_{min}$ can be found from the
following relations:
 \begin{equation}
 \renewcommand{\arraystretch}{1.8}
 \begin{array}{lll}
   d_{\rm min}=r /\sqrt{1+(V_r/V_t)^2},\\
   t_{\rm min}=r V_r/(V^2_t+V^2_r),
 \label{lin}
 \end{array}
 \end{equation}
where $V_t=\sqrt{V^2_l+V^2_b}$ is the stellar velocity perpendicular to the line of sight.

 \subsubsection*{\it Model Gravitational Potential}
The axisymmetric Galactic potential is represented as a sum of three components—a central
spherical bulge $\Phi_b(r(R,Z)),$ a disk$\Phi_d(r(R,Z)),$  and a massive spherical dark matter halo $\Phi_h(r(R,Z))$:
 \begin{equation}
 \begin{array}{lll}
  \Phi(R,Z)=\Phi_b(r(R,Z))+\Phi_d(r(R,Z))+\Phi_h(r(R,Z)).
 \label{pot}
 \end{array}
 \end{equation}
Here, we use a cylindrical coordinate system ($R,\psi,Z$) with the coordinate origin at the Galactic center. In a rectangular coordinate system $(X,Y,Z)$ the distance to a star (spherical radius) will be $r^2=X^2+Y^2+Z^2=R^2+Z^2.$ The gravitational potential is
expressed in units of 100 km$^2 s^{-2}$, the distances are in kpc, and the masses are in units of the Galactic mass $M_{\rm gal}=2.325\times 10^7 M_\odot$ corresponding to the gravitational constant $G=1.$

The bulge, $\Phi_b(r(R,Z)),$ and disk, $\Phi_d(r(R,Z)),$
potentials are represented in the form proposed by
Miyamoto and Nagai (1975):
 \begin{equation}
  \Phi_b(r)=-\frac{M_b}{(r^2+b_b^2)^{1/2}},
  \label{bulge}
 \end{equation}
 \begin{equation}
 \Phi_d(R,Z)=-\frac{M_d}{\Biggl[R^2+\Bigl(a_d+\sqrt{Z^2+b_d^2}\Bigr)^2\Biggr]^{1/2}},
 \label{disk}
\end{equation}
where $M_b$ and $M_d$ are the masses of the components, $b_b, a_d,$ and $b_d$ are the scale lengths of the components in kpc. The halo component is represented according
to Navarro et al. (1997):
 \begin{equation}
  \Phi_h(r)=-\frac{M_h}{r}\ln {\Biggl(1+\frac{r}{a_h}\Biggr)}.
 \label{halo-III}
 \end{equation}
The parameters of the model Galactic potential (4)--(6) are given in Table 1. In Bajkova and Bobylev (2016b) the model (4)--(6) is designated as model III.

%%%%%%%%%%%%%%%%%%%%%%%%%%%%%%%%%%%%%%%%%%%%%%%%%%%%%%%%
 {\begin{table}[t]                                    %% t1.
 \caption[]
 {\small\baselineskip=1.0ex
 Parameters of the model Galactic potential from Bajkova and Bobylev (2016b), 
 $M_{\rm gal}=2.325\times 10^7 M_\odot$
  }
 \label{t:model-III}
 \begin{center}\begin{tabular}{|c|c|r|}\hline
 Parameters           &  Model III\\\hline
 $M_b$($M_{\rm gal}$) &    443$\pm27$  \\
 $M_d$($M_{\rm gal}$) &   2798$\pm84$ \\
 $M_h$($M_{\rm gal}$) &  12474$\pm3289$ \\
 $b_b$(kpc)       & 0.2672$\pm0.0090$ \\
 $a_d$(kpc)       &   4.40$\pm0.73$ \\
 $b_d$(kpc)       & 0.3084$\pm0.0050$ \\
 $a_h$(kpc)       &    7.7$\pm2.1$ \\\hline
 \end{tabular}\end{center}\end{table}}
%%%%%%%%%%%%%%%%%%%%%%%%%%%%%%%%%%%%%%%%%%%%%%%%%%%%%%%%

The equations of motion for a test particle in a Galactic potential appear as follows:
\begin{equation}
 \begin{array}{llllll}
 \dot{X}=p_X,\quad
 \dot{Y}=p_Y,\quad
 \dot{Z}=p_Z,\\
 \dot{p}_X=-\partial\Phi/\partial X,\\
 \dot{p}_Y=-\partial\Phi/\partial Y,\\
 \dot{p}_Z=-\partial\Phi/\partial Z,
 \label{eq-motion}
 \end{array}
\end{equation}
where $p_X, p_Y,$ and $p_Z$ are the canonical momenta, the dot denotes a time derivative. The fourth-order Runge–Kutta algorithm was used to integrate Eqs. (7).

In the rectangular Galactic coordinate system the initial test particle positions and velocities are determined from the formulas
\begin{equation}
 \begin{array}{llllll}
 X=R_0-X_0, ~Y=Y_0, ~Z=Z_0+h_\odot,\\
 U=-(U_0+U_\odot),\\
 V=V_0+V_\odot+V_{circ},\\
 W=W_0+W_\odot,
 \label{init}
 \end{array}
\end{equation}
where $(X_0,Y_0,Z_0,U_0,V_0,W_0)$ are the initial test particle positions and space velocities in the heliocentric coordinate system and the circular rotation velocity
of the solar neighborhood in our potential is $V_{\rm circ}=244$~km s$^{-1}$. The peculiar velocity components of the Sun $V_{\rm circ}=244$~km s$^{-1}$ were taken from Sch\"onrich et al. (2010). The Sun’s height above the Galactic plane $h_\odot=16$ pc was taken from
Bobylev and Bajkova (2016a).

As previously, for each star we calculate the encounter parameter between the stellar and solar orbits $d(t)=\sqrt{\Delta X^2(t)+\Delta Y^2(t)+\Delta Z^2(t)}$. Then, we determine
$d_{min}$ at the encounter time $t_{min}.$

We estimate the errors in $d_{min}$ and $t_{min}$ by the Monte Carlo method. Here the errors in the stellar parameters are assumed to be distributed normally with a dispersion $\sigma$. The errors are added to the stellar equatorial coordinates, proper motion components,
parallaxes, and line-of-sight velocities.

 \subsubsection*{\it Inclusion of a Spiral DensityWave}
In the case where the spiral density wave is taken into account (Lin and Shu 1964; Lin et al. 1969), the following term (Fernandez et al. 2008) is added to the
right-hand side of Eq. (3):
\begin{equation}
 \Phi_{sp} (R,\theta,t)= A\cos[m(\Omega_p t-\theta)+\chi(R)].
 \label{Potent-spir}
\end{equation}
Here
 $$
 A= \frac{(R_0\Omega_0)^2 f_{r0} \tan i}{m},\qquad
 \chi(R)=- \frac{m}{\tan i} \ln\biggl(\frac{R}{R_0}\biggr)+\chi_\odot,
 $$
where $A$ is the amplitude of the spiral wave potential, $f_{r0}$ is the ratio between the radial component of the perturbation from the spiral arms and the Galaxy’s general attraction, $\Omega_p$ is the pattern speed, $m$ is the number of spiral arms, $i$ is the pitch angle of the arms ($i<0$ for a wound pattern), $\chi$ is the radial wave phase
(the arm center corresponds to $\chi=0^\circ$), and $\chi_\odot$ is the radial phase of the Sun in the spiral wave.

In this paper the following spiral wave parameters were taken as a first approximation:
 \begin{equation}
 \begin{array}{lll}
 m=4,\\
 i=-13^\circ,\\
 f_{r0}=0.05,\\
 \chi_\odot=-140^\circ,\\
 \Omega_p=20~\hbox {km s$^{-1}$ kpc$^{-1}.$}
 \label{param-spiral}
 \end{array}
 \end{equation}
This set of parameters was adopted in Bobylev and Bajkova (2014), where an overview of the publications on this subject can be found. Note that the model potential can be even more complex and contain the contribution of a central bar (see, e.g.,
Garcia-S\'anchez et al. 2001). However, we decided
to neglect the influence of the bar, because the characteristics
of the central bar in the Galaxy (according to some data, two bars) are known with an even
greater uncertainty in comparison with the spiral wave characteristics.

%%%%%%%%%%%%%%%%%%%%%%%%%%%%%%%%%%%%%%%%%%%%%%%%%
 \begin{table}[t]                               %t1
 \caption[]{\small\baselineskip=1.0ex\protect
 Input data on the stars
 }
 \begin{center}
 \begin{tabular}{|r|r|r|r|r|r|r|r|}\hline
 \label{tab-1}

 \def\baselinestretch{1}\normalsize\small
 Gaia\,EDR3 &  $\pi,$  & $\mu_\alpha\cos\delta,$ & $\mu_\delta,$ & $V_r,$ \\
            &      mas &  mas yr$^{-1}$          &  mas yr$^{-1}$  & km s$^{-1}$ \\\hline

4270814637616488064&$ 52.40\pm0.02$&$   -0.41\pm0.02$&$   -0.11\pm0.02$&$ -14.47\pm 0.02$ \\
 510911618569239040&$ 13.21\pm0.03$&$    0.14\pm0.02$&$    0.01\pm0.03$&$  26.45\pm 0.35$ \\
5571232118090082816&$ 10.23\pm0.01$&$    0.09\pm0.01$&$    0.46\pm0.01$&$  82.18\pm 0.47$ \\
 729885367894193280&$ 20.70\pm0.84$&$    0.64\pm0.96$&$   -2.35\pm1.31$&$ -90 \pm54 $ \\
1952802469918554368&$141.90\pm0.02$&$  161.45\pm0.02$&$ -119.74\pm0.02$&$ -98.52\pm 7.54$ \\
6396469681261213568&$  9.80\pm0.02$&$    0.49\pm0.01$&$    0.19\pm0.02$&$  52.30\pm 0.24$ \\
3118526069444386944&$  7.61\pm0.055$&$  0.252\pm0.05$&$   0.045\pm0.05$&$  40.41\pm0.94$ \\
1281410781322153216&$ 20.80\pm0.04$&$    1.29\pm0.03$&$    1.00\pm0.03$&$  31.84\pm4.73$ \\
1949388868571283200&$  4.15\pm0.02$&$   -0.47\pm0.02$&$   -0.63\pm0.02$&$ 347.3 \pm 6.5 $ \\
5261593808165974784&$ 15.36\pm0.01$&$   -0.09\pm0.01$&$   -2.21\pm0.02$&$  71.05\pm 0.88$ \\
2595284016771502080&$138.23\pm0.05$&$  308.71\pm0.05$&$ -718.39\pm0.04$&$  308  \pm116  $ \\
1251059445736205824&$ 24.37\pm0.21$&$   -0.23\pm0.20$&$   -3.24\pm0.16$&$  40   \pm10 $ \\
1227133699053734528&$107.73\pm0.22$&$   86.67\pm0.29$&$  127.95\pm0.20$&$ -87   \pm33 $ \\
1791617849154434688&$ 11.38\pm0.02$&$   -0.39\pm0.01$&$   -1.17\pm0.01$&$  56.29\pm 0.48$ \\
2926732831673735168&$  8.85\pm0.01$&$   -0.74\pm0.01$&$    0.53\pm0.01$&$  66.49\pm 0.25$ \\
3260079227925564160&$ 32.11\pm0.03$&$   -3.62\pm0.03$&$   -4.96\pm0.02$&$ -33.38\pm 0.42$ \\
3972130276695660288&$ 59.92\pm0.03$&$  -20.81\pm0.03$&$    6.63\pm0.02$&$  31.80\pm 0.73$ \\
1926461164913660160&$316.48\pm0.04$&$  112.53\pm0.04$&$-1591.65\pm0.03$&$ -78.00\pm 0.40$ \\
2552928187080872832&$231.78\pm0.02$&$ 1231.40\pm0.02$&$-2711.88\pm0.02$&$  263.0\pm 4.9 $ \\
1129149723913123456&$190.33\pm0.02$&$  748.42\pm0.02$&$  480.80\pm0.03$&$-111.65\pm0.02$ \\
2924378502398307840&$  6.10\pm0.01$&$    0.75\pm0.01$&$   0.14\pm0.01$& $  86.98\pm 1.00$ \\
6608946489396474752&$  7.93\pm0.01$&$   -0.65\pm0.01$&$  -0.31\pm0.01$& $  44.23\pm 0.57$ \\
   \hline
   \end{tabular} \end{center}
 \end{table}
%%%%%%%%%%%%%%
%%%%%%%%%%%%%%%%%%%%%%%%%%%%%%%%%%%%%%%%%%%%%%%%%
 \begin{table}[t]                               %t2
 \caption[]{\small\baselineskip=1.0ex\protect
 Additional data on the stars
 }
 \begin{center}
 \begin{tabular}{|r|r|c|c|c|}\hline
 \label{tab-2}

 \def\baselinestretch{1}\normalsize\small
        Gaia\,EDR3 &       Alternative name  & StePPeD & Mass, $M_\odot$ \\\hline
4270814637616488064&                  GJ 710 & P0107 & 0.650 \\
 510911618569239040&         TYC 4034-1077-1 & P0230 & 1.100 \\
5571232118090082816&                         & P0506 & 0.766 \\
 729885367894193280& 2MASS J10492824+2537231 & P0414 & 0.080 \\
1952802469918554368&                         & P0416 & 0.200 \\
6396469681261213568&          TYC 9327-264-1 & P0382 & 0.891 \\
3118526069444386944&                         & P0533 & 0.865 \\
1281410781322153216&              WD 1446+28 & P0417 & 0.852 \\
1949388868571283200&                         & P0524 & 0.695 \\
5261593808165974784&                         & P0522 & 0.547 \\
2595284016771502080&                 GJ 4274 & P0412 & 0.139 \\
1251059445736205824& 2MASS J13510178+2200085 & P0423 & 0.100 \\
1227133699053734528& 2MASS J14162408+1348263 & P0457 & 0.080 \\
1791617849154434688&         TYC 1662-1962-1 & P0189 & 0.710 \\
2926732831673735168&         TYC 5960-2077-1 & P0287 & 1.023 \\
3260079227925564160&                         & P0526 & 0.450 \\
3972130276695660288&                 GJ 3649 & P0178 & 0.549 \\
1926461164913660160&                  GJ 905 & P0413 & 0.151 \\
2552928187080872832&              WD 0046+05 & P0005 & 0.500 \\
1129149723913123456&               HIP 57544 & P0078 & 0.294 \\
2924378502398307840&          UCAC2 21925028 & P0400 & 0.709 \\
6608946489396474752&                         & P0514 & 0.746 \\
   \hline
   \end{tabular} \end{center}
 \end{table}
%%%%%%%%%%%%%%
%%%%%%%%%%%%%%%%%%%%%%%%%%%%%%%%%%%%%%%%%%
 \begin{table}[t]                               %T~3
 \caption[]{\small\baselineskip=1.0ex\protect
 Parameters of the stellar encounters with the Solar system
 }
 \begin{center}
 \begin{tabular}{|r|rr|rr|rr|rr|}\hline
 \label{tab-rez}
\def\baselinestretch{1}\normalsize\small
Gaia\,EDR3  & $t_{\rm min},$ & $d_{\rm min},$ & $t_{\rm min},$ & $d_{\rm min},$ & $t_{\rm min},$ & $d_{\rm min},$ & $\sigma_t,$ & $\sigma_d,$\\
                    &  Myr   &  pc  &   Myr   &  pc   &  Myr  &  pc  & Myr & pc \\\hline
                    &     (1)&       &      (2)&       &       (3)&       &       &      \\\hline
 4270814637616488064&$ 1.320$&$0.051$&$ 1.320$&$ 0.020$&$ 1.320$& 0.020& .028& .007 \\
  510911618569239040&$-2.861$&$0.149$&$-2.863$&$ 0.057$&$-2.863$& 0.066& .046& .079 \\
 5571232118090082816&$-1.189$&$0.259$&$-1.189$&$ 0.196$&$-1.189$& 0.190& .021& .021 \\
  729885367894193280&$ 0.537$&$0.300$&$ 0.538$&$ 0.300$&$ 0.538$& 0.300& 1.31& 1.92 \\%Dyb
 1952802469918554368&$ 0.071$&$0.479$&$ 0.072$&$ 0.462$&$ 0.072$& 0.462& .006& .039 \\%Dyb
 6396469681261213568&$-1.950$&$0.495$&$-1.946$&$ 0.867$&$-1.946$& 0.880& .011& .024 \\%Dyb
 3118526069444386944&$-3.253$&$0.521$&$-3.259$&$ 0.509$&$-3.262$& 0.525& .079& .097 \\%Dyb
 1281410781322153216&$-1.510$&$0.563$&$-1.507$&$ 0.499$&$-1.507$& 0.498& .747& .625 \\%Dyb
 1949388868571283200&$-0.693$&$0.622$&$-0.694$&$ 0.660$&$-0.694$& 0.657& .015& .134 \\
 5261593808165974784&$-0.917$&$0.626$&$-0.917$&$ 0.650$&$-0.917$& 0.650& .012& .014 \\
 2595284016771502080&$-0.023$&$0.627$&$-0.024$&$ 0.604$&$-0.024$& 0.604& .016& .505 \\%Dyb
 1251059445736205824&$-1.025$&$0.647$&$-1.024$&$ 0.603$&$-1.024$& 0.603& .417& .248 \\%Dyb
 1227133699053734528&$ 0.106$&$0.723$&$ 0.107$&$ 0.708$&$ 0.107$& 0.708& .10 &1.03 \\%Dyb
 1791617849154434688&$-1.560$&$0.802$&$-1.561$&$ 0.850$&$-1.561$& 0.843& .014& .040 \\%Dyb
 2926732831673735168&$-1.699$&$0.827$&$-1.700$&$ 0.794$&$-1.700$& 0.788& .007& .022 \\
 3260079227925564160&$ 0.932$&$0.845$&$ 0.933$&$ 0.784$&$ 0.934$& 0.783& .013& .011 \\
 3972130276695660288&$-0.523$&$0.906$&$-0.523$&$ 0.892$&$-0.523$& 0.892& .013& .024 \\%Dyb
 1926461164913660160&$ 0.037$&$0.926$&$ 0.037$&$ 0.909$&$ 0.037$& 0.909& .001& .004 \\
 2552928187080872832&$-0.016$&$0.973$&$-0.016$&$ 1.017$&$-0.016$& 1.017& .001& .019 \\
 1129149723913123456&$ 0.045$&$1.023$&$ 0.046$&$ 1.004$&$ 0.046$& 1.004& .000& .002 \\
 6726602067616477056&$-2.140$&$1.030$&$-2.143$&$ 0.965$&$-2.144$& 0.963& .004& .021 \\%Dyb
 2924378502398307840&$-1.885$&$1.114$&$-1.885$&$ 0.921$&$-1.885$& 0.921& .022& .059 \\
 6608946489396474752&$-2.849$&$1.228$&$-2.820$&$ 0.571$&$-2.821$& 0.552& .039& .034 \\
   \hline
   \end{tabular} \end{center}
 {\small (1) the linear method, (2) an axisymmetric potential, (3) a potential with a spiral density wave.}
 \end{table}
%%%%%%%%%%%%%

 \subsection*{DATA}
The working sample was produced as follows. First we compiled a preliminary list of stellar candidates for close encounters with the Solar system (with an encounter parameter less than 1 pc). The Stellar Potential Perturbers Database (StePPeD)~\footnote{https://pad2.astro.amu.edu.pl/stars} described by Wysocza\'nska et al. (2020) served as the main source for this purpose. Data from the Gaia DR2 catalogue were used to create this database. We added several stars from Bobylev and Bajkova (2020). About 50 stars were included in this preliminary list.

Then, we identified the stars from the preliminary list with the Gaia EDR3 catalogue. Unfortunately, there were no parallax measurements in the
new version of the Gaia catalogue for several stars
of interest in the search for encounters. For example,
such measurements are absent for the star
ALS 9243, which could approach the solar orbit to
a distance of 0.25 pc, as estimated by Wysocza\'nska
et al. (2020), 2.5 Myr ago. They are also absent for
the record-holder in encounters from StePPeD---the
star Gaia DR2 4535062706661799168. For some
stars (Gaia DR2 969867803725057920 or Gaia DR2
365942724131566208) their new parallaxes lead to such $d_{min}$ that delete these stars from the list of candidates for close encounters.

Such data on the selected stars as the name in the Gaia EDR3 catalogue, the parallax $\pi$, the propermotion components $\mu_\alpha\cos\delta$ and $\mu_\delta$, and the heliocentric
line-of-sight velocity $V_r$ are presented in Table 2. For these stars Table 3 gives an alternative name (if available), the name in StePPeD, and the mass estimate
copied from StePPeD.

For almost all of these stars their heliocentric line-of-sight velocities $V_r$ coincide with those given in StePPeD. However, there are exceptions. These include the white dwarfs WD 1446+28 and WD 0046+05.

For the white dwarf WD 1446+28 StePPeD gives $V_r=36.0\pm119.9$ km s$^{-1}$, which was measured with a very large error. In this paper for this star we took the heliocentric velocity $V_r=31.84\pm4.73$ km s$^{-1}$ from Anguiano et al. (2017), where the measurements
were performed much more accurately. Most importantly, the gravitational redshift was taken into account, which is relevant for white dwarfs, because, on average, this correction is $\sim$50 km s$^{-1}$ (Greenstein and Trimble 1967).

The white dwarf WD 0046+05 is also known as van Maanen’s star 2. There is an extensive bibliography where the spectroscopic observations of this star are described (Greenstein and Trimble 1967; Greenstein 1972; Gatewood and Russell 1974). According to these authors, the heliocentric velocity of the white dwarf WD 0046+05 is close to $V_r\sim1\pm15$ km s$^{-1}$
calculated by applying a correction for the gravitational redshift.

 \subsection*{RESULTS AND DISCUSSION}
Table 4 gives the parameters of the stellar encounters with the Solar system derived by three methods: the linear one (2), by integrating the orbits in an axisymmetric
potential (3), an by integrating the orbits in a potential with a spiral density wave (9). The last column gives the errors in the parameters that can be attributed to all three methods. These errors were estimated by the Monte Carlo method.

As can be seen from Table 4, the encounter parameters derived by the second and third methods barely differ. The encounter times $t_{min}$ found by all three methods are in excellent agreement between themselves: the discrepancy typically does not exceed
1--2 units of the second decimal place. In contrast, the difference in the distances dmin found by the first and second methods can reach 0.6~pc (for example, for
the star Gaia EDR3 6608946489396474752), though this difference is usually much smaller.

Quite a few stars from Table 4 were analyzed in Bobylev and Bajkova (2020) using data from the Gaia DR2 catalogue (note that the specific digital numbers of our stars in the DR2 and EDR3 versions coincide). We may conclude that the random errors in the encounter parameters $\sigma_t$ and $\sigma_d$ found in this paper decreased approximately by 30\% compared to the results of our analysis of the Gaia DR2 data. At the same time, there are two stars with huge random measurement errors of the line-of-sight velocities (see Table 2), which determine the huge (exceeding 1 pc in distance) errors $\sigma_t$ and $\sigma_d$. These
are the stars Gaia EDR3 729885367894193280 and Gaia EDR3 1227133699053734528.

As our calculations showed, with the heliocentric velocity of the white dwarf WD 0046+05 $V_r\sim1\pm15$ km s$^{-1}$ any close encounters of this star with the
Solar system are ruled out.

The star GJ 710 (the first row in Tables 2--4), which is known as one of the record-holders in very close encounters, is of great interest. For example, Bobylev and Bajkova (2020) derived the followin encounter parameters for it with the data from the
Gaia DR2 catalogue: $t_{min}=1.316\pm0.040$~Myr, $d_{min}=0.055\pm0.009$~pc using the linear method (method 1) and $t_{min}=1.320\pm0.040$ Myr, $d_{min}=0.016\pm0.009$~pc
by integrating the orbits in an axisymmetric potential (method 2). We see that using
the data from the Gaia EDR3 catalogue here led only
to a decrease in the random errors $\sigma_t$ and $\sigma_d$.

There are also examples of a significant change in the encounter parameters $t_{min}$ and $d_{min},$ found by using the data from the Gaia EDR3 catalogue. For example, for the star Gaia EDR3 3118526069444386944 with the data from the Gaia DR2 catalogue Wysocza\'nska et al. (2020) obtained the following encounter parameters by method 2: $t_{min}=-3.235$~Myr
and $d_{min}=0.979$~pc. As can be seen from Table 4, we found $t_{min}=-3.259\pm0.079$~Myr and $d_{min}=0.509\pm0.097$~pc by a similar method. Here using the latest measurements led to a significant decrease in the parameter $d_{min}.$ The star became more interesting for our problem, because it could pass along the edge of the Oort cloud.

This is also true for the star Gaia EDR3 510911618569239040, for which the
encounter parameter dmin decreased significantly.
Now it occupies the second row in our tables.
Wysocza\'nska et al. (2020) obtained the following
encounter parameters for this star by method 2:
$t_{min}=-2.789$~Myr and $d_{min}=0.412$~pc.

As a result, we can select the following five stars:
Gaia EDR3 4270814637616488064 (GJ 710),
Gaia EDR3 510911618569239040,
Gaia EDR3 5571232118090082816,
Gaia EDR3 1952802469918554368,
and Gaia EDR3 3118526069444386944. Applying any of the three methods shows that they are good candidates for penetration into the Oort cloud. In this list we did not include two stars with large errors $\sigma_t$ and $\sigma_d$. The star Gaia EDR3 6396469681261213568,
for which there are noticeable discrepancies in estimating the parameter $d_{min}$ by various methods, did not enter into this list either.

 \subsection*{CONCLUSIONS}
We considered a sample of 23 candidates for close (within 1 pc) encounters with the Solar system. The trigonometric parallaxes and proper motions of these stars were taken from the latest Gaia EDR3 catalogue. The stellar encounter parameters were calculated using the linear method (1), by integrating the orbits in an axisymmetric potential (2), and by integrating the orbits in a potential with a spiral density
wave (3). We concluded that the results obtained
by the second and third methods barely differ. The
encounter parameters derived by the first method are
in good agreement with the results obtained by the
other two methods, although the difference in the
distances $d_{min}$ found by the first and the other two
methods can reach several tenths of a parsec in some cases.

 \bigskip \bigskip\medskip{\bf REFERENCES}{\small

1. E. Anderson and Ch. Francis, Astron. Lett. 38, 331 (2012).

2. B. Anguiano, A. Rebassa-Mansergas, E. Garcia-Berro, S. Torres, K. C. Freeman, and T. Zwitter, Mon. Not. R. Astron. Soc. 469, 2102 (2017).

3. P. Bacci, M. Maestripieri, L. Tesi, G. Fagioli, R. A. Mastaler, G. Hug, M. Schwartz,
R. R. Holvorcem, et al., Minor Planet Electron. Circ., no. 2017--U181 (2017).

4. C. A. L. Bailer-Jones, Astron. Astrophys. 575, 35 (2015).

5. C. A. L. Bailer-Jones, Astron. Astrophys. 609, 8 (2018).

6. C. A. L. Bailer-Jones, J. Rybizki, R. Andrae, and M. Fouesneau, Astron. Astrophys. 616, 37 (2018).

7. F. Berski and P. A. Dybczy\'nski, Astron. Astrophys. 595, L10 (2016).

8. V. V. Bobylev, Astron. Lett. 36, 220 (2010a).

9. V. V. Bobylev, Astron. Lett. 36, 816 (2010b).

10. V. V. Bobylev and A. T. Bajkova, Astron. Lett. 40, 352 (2014).

11. V. V. Bobylev, and A. T. Bajkova, Astron. Lett. 42, 1 (2016a).

12. V. V. Bobylev and A. T. Bajkova, Astron. Lett. 42, 567 (2016b).

13. V. V. Bobylev and A. T. Bajkova, Astron. Lett. 43, 559 (2017).

14. V. V. Bobylev and A. T. Bajkova, Astron. Lett. 46, 245 (2020).

15. G. Borisov, Minor Planet Electron. Circ., no. 2019--R106, 11 (2019).

16. A. G. A. Brown, A. Vallenari, T. Prusti, de Bruijne, C. Babusiaux, C. A. L. Bailer-Jones, M. Biermann, D. W. Evans, et al. (Gaia Collab.), Astron. Astrophys.
616, 1 (2018).

17. A. G. A. Brown, A. Vallenari, T. Prusti, J. H. J. de Bruijne, C. Babusiaux, M. Biermann,
O. L. Creevey, D. W. Evans, et al. (Gaia Collab.), arXiv: 2012.01533 (2020).

18. R. Darma, W. Hidayat, and M. I. Arifyanto, J. Phys.: Conf. Ser. 1245, 012028 (2019).

19. P. A. Dybczy\'nski, Astron. Astrophys. 396, 283 (2002).

20. P. A. Dybczy\'nski, Astron. Astrophys. 441, 783 (2005).

21. P. A. Dybczy\'nski and F. Berski, Mon. Not. R. Astron. Soc. 449, 2459 (2015).

22. ESA, The Hipparcos and Tycho Catalogues, ESA SP--1200 (ESA, 1997).

23. F. Feng and C. A. L. Bailer-Jones, Mon. Not. R. Astron. Soc. 454, 3267 (2015).

24. D. Fernandez, F. Figueras, and J. Torra, Astron. Astrophys. 480, 735 (2008).

25. R. de la FuenteMarcos and C. de la Fuente Marcos, Res. Not. Am. Astron. Soc. 2, 30 (2018).

26. J. Garcia-S\'anchez, R. A. Preston, D. L. Jones, P. R. Weissman, J.-F. Lestrade, D. W. Latham, and R. P. Stefanik, Astron. J. 117, 1042 (1999).

27. J. Garcia-S\'anchez, P. R. Weissman, R. A. Preston, D. L. Jones, J.-F. Lestrade, D.W. Latham, R. P. Stefanik, and J. M. Paredes, Astron. Astrophys. 379, 634 (2001).

28. G. Gatewood and J. Russell, Astron. J. 79, 815 (1974).

29. J. L. Greenstein and V. L. Trimble, Astrophys. J. 149, 283 (1967).

30. J. L. Greenstein, Astrophys. J. 173, 377 (1972).

31. C. C. Lin and F. H. Shu, Astrophys. J. 140, 646 (1964).

32. C. C. Lin, C. Yuan and F. H. Shu, Astrophys. J. 155, 721 (1969).

33. L. Lindegren, U. Lammers, U. Bastian, J. Hernandez, S. Klioner, D. Hobbs, A. Bombrun, D. Michalik, et al. (Gaia Collab.), Astron. Astrophys. 595, A4 (2016).

34. L. Lindegren, J. Hern\'andez, A. Bombrun, S. Klioner, U. Bastian, M. Ramos-Lerate, A. de Torres, H. Steidelm\"uller, et al. (Gaia Collab.), Astron. Astrophys. 616, 2 (2018).

35. L. Lindegren, S. A. Klioner, J. Hern\'andez, A. Bombrun, M. Ramos-Lerate, H. Steidelm\"uller, U. Bastian, M. Biermann, et al. (Gaia Collab.), arXiv: 2012.03380 (2020).

36. T. E. Lutz and D. H. Kelker, Publ. Astron. Soc. Pacif. 85, 573 (1973).

37. C. A. Martinez-Barbosa, L. J\'ylkov\'a, S. Portegies Zwart,
and A.G. A. Brown, Mon. Not. R. Astron. Soc. 464, 2290 (2017).

38. R. A. J. Matthews, R. Astron. Soc. Quart. J. 35, 1 (1994).

39. M. Miyamoto and R. Nagai, Publ. Astron. Soc. Pacif. 27, 533 (1975).

40. A. A. M\"ull\"ari and V. V. Orlov, Earth, Moon, and Planets (Kluwer, Netherlands, 1996), Vol. 72, p. 19.

41. J. F. Navarro, C. S. Frenk, and S. D. M. White, Astrophys. J. 490, 493 (1997).

42. J. H. Oort, Bull. Astron. Inst. Netherl. 11 (408), 91
(1950).

43. S. Portegies Zwart, arXiv: 2011.08257 (2020).

44. I. A. Revina, Analysis of the Motion of Celestial Bodies and Estimation of the Accuracy of their Observations (Latvian University, Riga, 1988), p. 121 [in Russian].

45. R. Sch\"onrich, J. Binney, and W. Dehnen, Mon. Not. R. Astron. Soc. 403, 1829 (2010).

46. S. Torres, M. X. Cai, A. G. A. Brown, and S. Portegies Zwart, Astron. Astrophys. 629, 139 (2019).

47. R. Wysocza\'nska, P. A. Dybczy\'nski, and M. Poli\'nska, arXiv: 2003.02069 (2020).
  }
  \end{document}